\documentclass[namedreferences]{SolarPhysics}
\usepackage[optionalrh]{spr-sola-addons} 
\usepackage{graphicx}                    
\usepackage{color}                       
\usepackage{url}                         

\newcommand{\aap}{    {\it Astron. Astrophys.}}

\newcommand{\apj}{    {\it Astrophys. J.}}

\newcommand{\mnras}{  {\it Mon. Not. Roy. Astron. Soc.}}

\newcommand{\pasj}{   {\it Publ. Astron. Soc. Japan}}

\newcommand{\solphys}{{\it Solar Phys.}}
\newcommand{\sovast}{ {\it Sov. Astron.}}

\newcommand{\kfnt}{   {\it Kinematika Fiz. Nebesnykh. Tel}}
\newcommand{\kpcb}{   {\it Kin. Phys. Cel. Bodies}}
\newcommand{\an}{  {\it Astron. Nachr.}}
\newcommand{\apjs}{    {\it Astrophys. J. Suppl.}}

\begin{document}

\begin{article}
\begin{opening}

\title{Center--limb variation of solar photospheric microturbulence}

%
\author{Yoichi~\surname{Takeda}}

%
\runningauthor{Y. Takeda}
\runningtitle{Center--limb variation of microturbulence}

%
  \institute{11-2 Enomachi, Naka-ku, Hiroshima-shi 730-0851, Japan\\
                     email: \url{ytakeda@js2.so-net.ne.jp}\\ 
             }

\begin{abstract}
Microturbulence ($\xi$) is a key parameter introduced in 
stellar spectroscopy to explain the strength of saturated lines 
by formally incorporating an additional thermal broadening 
term in the line opacity profile. Although our Sun can serve as an 
important testing bench to check the usual assumption of constant 
$\xi$, the detailed behavior of how $\xi$ varies from the disk center 
through the limb seems to have never been investigated so far.
In order to fill this gap, local $\xi$ values on the solar disk were 
determined from the equivalent widths of 46 Fe~{\sc i} lines at 
32 points from the center to the limb by requiring the 
consistency between the abundances derived from lines of various strengths. 
The run of $\xi$ with $\theta$ (angle between line of sight and 
the surface normal) was found to be only gradual from 
$\approx$~1.0~km~s$^{-1}$ (at $\sin\theta = 0$: disk center) 
to $\approx$~1.3~km~s$^{-1}$ (at $\sin\theta \approx  0.7$: 
two-thirds of radial distance); but thereafter increasing more steeply 
up to $\approx$~2~km~s$^{-1}$ (at $\sin\theta = 0.97$: limb).  
This result further suggests that the microturbulence derived from the 
flux spectrum of the disk-integrated Sun is by $\approx 20$\% larger than
that of the disk-center value, which is almost consistent with the
prediction from 3D hydrodynamical model atmospheres.   
\end{abstract}

%
\keywords{Center-Limb Observations; 
Spectral Line, Broadening;
Spectrum, Visible;
Turbulence;
Velocity Fields, Photosphere}

\end{opening}

%

\section{Introduction}

It has been common in traditional stellar spectroscopy to 
invoke the very rough concept of micro/macro dichotomy in 
treating the effect of atmospheric turbulent velocity 
field on spectral lines: The ``micro''-turbulence with 
its characteristic scale-length considerably smaller
than the photon mean-free-path acts on the line {\it opacity}
to broaden as in the case of the thermal motion of atoms, 
while conversely the ``macro''-turbulence is a kind of global 
atmospheric motion which widens the emergent line {\it profile} 
similar to the case of rotational line broadening.

Especially, the former microturbulence (designated as $\xi$
in this paper, which is incorporated in the thermal Doppler velocity
term of line opacity profile as $\sqrt{2kT/m + \xi^{2}}$: 
$k$ is Boltzmann constant, $T$ is temperature, and $m$ is 
atomic mass) plays a significant role in stellar chemical 
abundance determination as an indispensable fudge parameter, 
which especially affects the lines of moderately large strengths 
on the shoulder or flat part of the curve of growth. 
Since the effect of this parameter depends on line strengths, 
it is usually determined by requiring that the abundances 
derived from lines of diversified equivalent widths are 
consistent with each other, which means that measurement 
for a number of lines of the same species ({\it e.g.}, Fe~{\sc i}) 
is necessary.

Alternatively, another (rather special) method is applicable, 
which makes use of the fact that the core profile of saturated 
lines is critically influenced by $\xi$ ({\it i.e.}, boxy shape
becomes more manifest with an increase in $\xi$).
In this case, Fourier transform technique is often employed,
because such a subtle difference of line profile is more easily
detected in the Fourier space ({\it e.g.}, position of zero amplitude
frequency). Although $\xi$ is in principle determinable from 
the profile of only one strong line, this approach is not 
so widely used in practice, since the results often show some
inconsistencies with those from the conventional method.     

Regarding the solar photospheric microturbulence, quite a few 
number of studies have been published so far, as shown 
in Table~1 which summarizes the data after 1976 (see also Table~VI 
of Blackwell {\it et al.}, 1976, where older references in 1955--1972 
are compiled). The following characteristics are read from 
these literature data of $\xi$ for the Sun.
\begin{itemize}
\item
Most of the results for solar $\xi$ are around $\approx 1$~km~s$^{-1}$ 
(within dispersion of $\approx \pm 0.5$km~s$^{-1}$). As a matter of 
fact, $\xi = 1$~km~s$^{-1}$ is an often adopted assumption for
the solar microturbulence ({\it cf.} Section~3.2 in Takeda, 1994).
\item
The microturbulence determined from line profile analysis appears
somewhat lower than that derived by the conventional method 
requiring the consistency of abundances.
\item
The $\xi$ values near to the limb tend to be larger than
that those at the disk center.
\end{itemize}  

The last point may be significant in context of conventional
spectroscopic analysis of stars in general (unresolved point 
source) where a single-valued $\xi$ has to be assumed. 
If microturbulence is variable over the disk, what kind of 
effect is expected on the result? 
As the only star whose surface is observable in detail, our 
Sun provides us with an opportunity to check this problem 
by clarifying the behavior of $\xi$ from the disk center 
through the limb. 
Unfortunately, to the author's knowledge, this has never been 
seriously investigated so far. What has been reported is only 
the fact of variable $\xi$ in a few sparse points at most ({\it e.g.}, 
Blackwell {\it et al.}, 1976; Holweger, Gehlsen, and Ruland, 1978; Gadun and Kostyk, 1990).
This is presumably due to the necessity of equivalent widths 
measurements for a large number of lines to establish $\xi$, 
and repeating this process for many different points on the disk 
may have been felt reluctant.
 
Recently, Takeda and UeNo (2019; hereinafter refereed to as TU19)
published the extensive equivalent width data set of 565 lines 
measured for 32 points (densely distributed from the solar disk 
center to the limb), which are just suitable for the task mentioned above. 
Motivated by this situation, the author decided to investigate the nature 
of center--limb variation in $\xi$ by making use of the data of TU19,
and to examine how it affects the conventional procedure of using 
constant $\xi$ in an application to disk-integrated spectrum. 
The aim of this article is to describe the outcome of this analysis.
 
\section{Observational Data}

Regarding the basic data of equivalent widths ($W$), those for 565 lines 
measured from the intensity spectra observed (by using the Domeless Solar 
Telescope at Hida Observatory of Kyoto University) at each of the 32 
different positions on the solar disk were used, which were 
published by TU19 as supplementary materials ({\it cf.} Appendix A.3 therein).

Since the use of spectral lines with sufficiently reliable $gf$ 
values is essential for the present purpose, the list of 65 Fe~{\sc i} 
lines adopted by Grevesse and Sauval (1999) for their solar Fe abundance 
analysis was invoked in this study. A comparison of these 65 lines
with those 565 lines in TU19 revealed that 50 lines are in common.
Then, after close inspection on these candidate lines, 4 lines 
turned out to show rather irregular and inadequate center--limb 
variation of $W$ and were thus discarded.
Consequently, 46 Fe~{\sc i} lines were decided to use as listed 
in Table~2, where the atomic parameters along with the equivalent 
widths at the disk center ($W_{00}$) and the limb ($W_{31}$) are
also presented.\footnote{Complete data of $W_{ii}$ ($ii = 00, \ldots, 31$) 
at 32 points between the center and the limb for these 46 lines are 
available in the supplementary material of TU19 (``all\_clvdata'').}  
In order for a comparison purpose, flux equivalent widths for the
disk-integrated Sun ($W_{\rm flux}$) were also measured for these 46 lines
based on Kurucz {\it et al.}'s (1984) solar flux spectrum by applying the
spectrum-fitting technique similar to that adopted in TU19
({\it cf.} Section~3.1 therein; the only difference is that flux is relevant 
in this case instead of specific intensity).

Comparison of disk-center equivalent widths derived in TU19 ($W_{00}$)
with those of Grevesse and Sauval (1999) ($W_{\rm d.c.}^{\rm GS}$)
is shown in Figure~1a, which shows that the latter tends to be 
systematically larger than the former by $\approx$~3\%.
Meanwhile, the $W_{\rm flux}$ values newly derived in this study are 
favorably compared with those of Takeda {\it et al.} (2005) (measured by 
the direct Gaussian fitting method), as displayed in Figure~1b.

Generally, Fe~{\sc i} lines tend to be strengthened with a decrease 
in temperature because of the negative sign of their $T$-sensitivity 
parameter ($K_{00}$ given in Table~2), 
\footnote{
This parameter is indicative of the $T$-sensitivity of $W_{00}$ at the 
disk center and defined as $K_{00} \equiv (d\log W/d\log T)_{00}$,
which was numerically evaluated for each line ({\it cf.} Equation~(6)
in TU19). 
}
which means that $W$ increases 
toward the limb ({\it i.e.}, towards shallower line-forming layer with lowered $T$). 
Therefore, $W_{31}/W_{00}$ is greater than unity and this tendency becomes 
more manifest towards a lesser $K_{00}$ ({\it i.e.}, larger $|K_{00}|$) 
as can be seen in Figure~1c, where the natural inequality of 
$W_{00} < W_{\rm flux} < W_{31}$ is also confirmed.   

Figure~1d shows the empirical curve of growth constructed from 
the disk center equivalent widths ($W_{00}$), which suggests that 
the adopted 46 Fe~{\sc i} lines are in the range of linear 
through flat part of the curve of growth. 

\section{Determination of Microturbulence} 

\subsection{Procedures}

As usually done, the value of $\xi$ at each disk point was determined 
by requiring that the abundances derived from each of the 46 lines 
do not show any systematic dependence upon $W$.
As in TU19, Kurucz's (1993) ATLAS9 model was adopted as the basic 
solar photospheric model, and likewise Kurucz's (1993) WIDTH9 
program (though modified in various respects) was used for
abundance calculation.
Regarding the atomic parameters, Grevesse and Sauval's (1999) paper 
was consulted for $\chi_{\rm low}$ and $\log gf$ ({\it cf.} Table~2),
while the adopted damping parameters are the same as given 
in ``tableE.dat'' of TU19.

Practically, the procedure employed by Blackwell {\it et al.} (1976)  
was followed in this investigation.\\
-- 1. For each line $n$, a set of abundances ($\log \epsilon_{n}^{m}$;  
$m = 1, 2,\ldots, M$) were derived from $W_{n}$ while incrementally 
changing the microturbulence ($\xi^{m}$; $m = 1, 2,\ldots, M$), 
where $\xi$ was varied from 0.0 to 2.6~km~s$^{-1}$ by an increment
of 0.01~km~s$^{-1}$ ({\it i.e.}, $M = 261$).\\ 
-- 2. Then, the mean abundance ($\langle \log\epsilon\rangle^{m}$)
averaged over $N (=46)$ lines and the standard deviation $\sigma^{m}$ 
were derived for each of the $M$ microturbulences ($\xi^{m}$).\\
-- 3. By inspecting the resulting set of ($\sigma^{m}$, $m = 1, 2,\ldots, M$),
the location of minimum $\sigma$  corresponds to desired 
solution, which is denoted by the subscript ``0'' such as $\xi_{0}$ 
(and likewise $\sigma_{0}$ or $\langle \log\epsilon\rangle_{0}$.\\

Such as done by Blackwell {\it et al.} (1976) in their Figs. 1 and 2,
the $\log\epsilon$ {\it vs.} $\xi$ relations for each of the 46 lines
along with the resulting $\sigma$ {\it vs.} $\xi$ curve are illustrated
in Figure~2 for the representative 6 points on the disk (panels a--f),
where the results for the disk center based on Grevesse and Sauval's 
(1999) $W_{\rm d.c.}$ (panel g) as well as for the disk-integrated Sun 
based on $W_{\rm flux}$ (panel h) are also presented.

\subsection{Error estimation}

While the mean error of $\langle\log\epsilon\rangle_{0}$ is
naturally expressed as $\sigma_{0}/\sqrt{N}$, 
it is not straightforward to estimate how much error is involved
in $\xi_{0}$. In this study, it was derived as follows.\\
-- Since the abundance for each line has an uncertainty of $\sigma$, 
randomly-generated noises ($\Delta$) of normal distribution (corresponding 
to a dispersion of $\sigma_{0}$) were added to the set of actual 
abundances as ($\log\epsilon_{n} + \Delta_{n}$, $n = 1, 2,\ldots, N$), 
from which a new standard deviation ($\sigma'$) was computed
and this process was repeated 1000 times ($k=1, 2, \ldots, 1000$).\\
-- Next, the standard deviation evaluated on a data set of 
($\sigma_{k}'/\sqrt{2}$, $k=1, 2, \ldots, 1000$) was denoted as
$\delta\sigma$,\footnote{Note that $\sigma'$ should be divided by 
$\sqrt{2}$ in order to avoid duplication, because the original 
abundance set already has an intrinsic dispersion of $\sigma_{0}$,}
which may be regarded as the ambiguity in the $\sigma (\xi)$ curve 
around the minimum.\\
-- Finally, the error ($\delta\xi$) involved in $\xi_{0}$ was estimated 
from the intersection points of the $\sigma(\xi)$ curve and the 
$\sigma = \sigma_{0} + \delta\sigma$ line, as they correspond 
to $\xi_{0} \pm \delta\xi$ (see Figure~2i, where this situation 
is graphically depicted). The resulting $\delta\xi$ turned out
typically on the order of $\approx 0.2$~km~s$^{-1}$. 

\section{Discussion}

\subsection{Resulting trends}

The finally established values of $\langle \log\epsilon\rangle_{0}$, 
$\sigma_{0}$, $\xi_{0}$, and $\delta\xi$ at each of the 32 points 
on the disk are given in Table~3, where the results derived from 
$W_{\rm d.c.}^{\rm GS}$ as well as $W_{\rm flux}$ are also presented
for comparison. In addition, the runs of these quantities from the
disk center to the limb are graphically illustrated in Figure~3.
The following characteristics are read from these results.
\begin{itemize}
\item
The $\langle \log\epsilon\rangle_{0}$ values at different disk 
points are remarkably similar (mean over the 32 points is 7.411 
with a standard deviation of 0.005). Likewise, the $\sigma_{0}$ 
values do not vary significantly from the center ($\approx 0.07$) 
to the limb ($\approx 0.08$). This implies that the classical
line formation theory using the microturbulence is practically 
validated irrespective of observed points on the disk (as far as
abundance determination is concerned). 
\item 
While a systematically increasing trend is surely observed in $\xi_{0}$ 
towards the limb, its rate is position-dependent in the sense that
a steeper gradient is seen near to the limb. Actually, the growth of
$\xi_{0}$ is only gradual at first from $\approx$~1.0~km~s$^{-1}$ 
(disk center) to $\approx$~1.3~km~s$^{-1}$ (at $\sin\theta \approx  0.7$: 
two-thirds of the radial distance); but it begins thereafter to 
increase more steeply until reaching $\approx$~2~km~s$^{-1}$ 
at the limb ($\sin\theta = 0.97$).  
\end{itemize}

Some previous authors have determined $\xi$ values at different points 
on the solar disk ({\it e.g.}, Blackwell {\it et al.}, 1976; Holweger, Gehlsen, and Ruland,  
1978; Gadun and Kostyk, 1990) as summarized in Table~1.
Since the observations of these literature studies were restricted 
to only a few points (usually at the disk center and the limb),
they are too coarse to be compared with the $\xi$ {\it vs.} $\theta$
trend derived in this investigation, Nevertheless, our $\xi$ 
values are more or less consistent with these results (within error 
bars) showing the same tendency of $\xi$(center)~$<$~$\xi$(limb).

\subsection{Reference formula for $\xi$}

It is useful to derive an analytical expression for the $\theta$-dependent 
$\xi$ as a reference model in preparation for the test calculations
presented in the following sections.

For this purpose, a slight pre-adjustment was applied to the original 
data of $\xi_{0}$ {\it vs.} $\theta$ relation. Although its disk-center 
value is 1.13~km~s$^{-1}$, the standard value of 1.0~km~s$^{-1}$ 
had better be adopted in order to maintain consistency with TU19,
where this value was used in deriving $K_{00}$ or $\log\epsilon_{00}$.
Therefore, $\xi_{0}(\theta)$ was divided by 1.13 to obtain the 
normalized $\xi_{\rm n}(\theta)$ to be employed, which is
a practically valid procedure given that $\xi_{0}$ is already 
uncertain by $\pm 0.2$~km~s$^{-1}$ (at any rate, our main concern 
is not so much the absolute values as the relative variation 
of $\xi$ on the disk). 

The resulting $\xi_{\rm n}(\theta)$ values are presented in 
the 8th column of Table~3. By applying a least-squares fitting 
with second-order polynomials (in terms of $\cos\theta$) to these data, 
the following analytical expression was found to yield a satisfactory fit
\begin{equation}
\xi_{\rm n} ({\rm km~s}^{-1}) = 1 + 0.0649 (1- \cos\theta) + 1.427 (1 - \cos\theta)^{2}
\end{equation} 
as demonstrated in Figures~4a and 4b.

It is worth mentioning that the $\xi_{\rm n}$ {\it vs.} $\theta$ relation 
expressed in terms of $\cos\theta$ shows a rather monotonically increasing 
trend towards the limb (Figure~4a), whereas the rate of increase becomes 
appreciably steeper near to the limb if $\sin\theta$ is used (Figure~4b).
This difference stems from the fact that, while specifying the disk position 
by $\sin\theta$ is in accord with the apparent view of the solar disk, 
the scale is more expanded near to the limb in case of using $\cos\theta$. 

\subsection{Consistency check}

It is worthwhile to confirm the validity of such obtained reference 
$\xi_{\rm n}(\theta)$ relation by checking that it does not cause
any inconsistency with other available information.

The first test is to find a empirical $\xi(\theta)$ relation
by demanding that the abundances derived from $W(\theta)$
do not depend upon $\theta$. In this method, the run of $\xi$ with 
$\theta$ can be evaluated from any single spectral line of 
sufficient strength, though $\xi$ at some fiducial point ({\it e.g.}, 
disk center) should be given in advance.   
That is, the disk center abundance $\log\epsilon_{00}$ was first 
derived from $W_{00}$ by assuming $\xi = 1$~km~s$^{-1}$; and the 
values of $\xi$ for the remaining points ($ii = 01, 02, \ldots, 31$) 
were determined from $W_{ii}$ by requiring that the abundance 
($\log\epsilon_{ii}$) be equal to $\log\epsilon_{00}$.
This procedure was applied to 15 lines (selected out of 46 lines) of 
medium-to-large strengths ($W_{00} \ge 50$~m\AA) and the resulting 
$\xi(\theta)$ relations are overplotted in Figures~4c and 4d,
where the curve expressed by Equation~(1) is also depicted for comparison.
We can see from these figures that the resulting trends (at least 
in the relative sense) are in reasonable agreement with each other.
 
The second check is to revisit the task once addressed in Appendix~1
of Takeda (2019), who  examined the mutual consistency of solar Fe abundances
derived at various points on the disk by adopting a constant microturbulence 
of $\xi = 1$~km~s$^{-1}$. In Figures~10a--10d of that paper, the abundance 
differences relative to the disk-center value are plotted against $\cos \theta$, 
which were derived for 280 Fe~{\sc i} lines (divided into 4 groups according to
the line strengths) based on $W$ values published by TU19. Those figures 
evidently indicated the existence of systematic discrepancies increasing
toward the limb. Then, what would happen if $\theta$-dependent $\xi$ given 
by Equation~(1) is used instead of $\xi = 1$~km~s$^{-1}$? The answers are shown 
in Figures~5a--5d, which are arranged in the same manner as the case of Takeda's 
(2019) Figures~10a--10d for the sake of enabling a direct comparison.
These figures suggest that the use of $\xi(\theta)$ has considerably removed 
the serious inconsistencies (seen in the case of constant $\xi$) to a 
satisfactory level, which may be counted as another evidence for the validity 
of the result in this study.\footnote{In Takeda (2019), an analytical formula of
$\xi = 1 + 0.6\sin\theta$ (km~s$^{-1}$) was applied as the $\theta$-dependent
solar microturbulence, which was tentatively devised in analogy with the 
$\theta$-dependence of the macroturbulence. That relation was not appropriate 
as viewed from the present knowledge, since it does not correctly reproduce 
the characteristic trend in terms of the increasing rate of $\xi$ (becoming steeper 
towards the limb).} 

\subsection{Impact on flux spectrum analysis}

Now that how the microturbulence varies over the solar disk has been
established, it is interesting to examine how such a position-dependent 
$\xi$ affects the spectrum analysis of the Sun-as-a-star 
where a constant $\xi$ is usually assumed.
For this purpose, the theoretical profiles (and the corresponding equivalent 
widths) of spectral lines were simulated by using $\theta$-dependent $\xi$, 
which were further analyzed inversely in the conventional manner 
by assuming a constant microturbulence. See Section~3.5 in Takeda (2019)
for the adopted program code of line profile simulation.

First, the examination was done on the usual microturbulence determination 
method. After the $W_{\rm cal,flux}$ values (theoretical equivalent width 
for the disk-integrated Sun) were calculated for 46 Fe~{\sc i} lines 
by assuming $\log\epsilon = 7.43$ and $\xi_{\rm n}(\theta)$ (Equation~(1)),
the critical $\xi^{*}$ value was derived for each line that yields an 
abundance of 7.43 by the classical analysis on $W_{\rm cal,flux}$.
The $\xi^{*}$ {\it vs.} $W_{\rm cal,flux}$ relations are depicted in Figure~6a,
where the plots are limited to lines of $W_{\rm cal,flux} > 10$~m\AA\ 
because too weak lines are insensitive to $\xi$. 
Besides, the abundances ($\log\epsilon$) were derived from $W_{\rm cal,flux}$ 
values of 46 lines for three cases of microturbulence (1.1, 1.2, and 
1.3~km~s$^{-1}$). The resulting $\log\epsilon$ {\it vs.} $W_{\rm cal,flux}$
plots are shown in Figure~6b.
These figures suggest that a conventional $\xi$-determination based on 
the $W_{\rm cal,flux}$ data leads to a result of $\xi \approx 1.2$~km~s$^{-1}$. 
  
Next, the influence on $\xi$-determination from the profile of a saturated line was 
investigated, which is often done by the location of first zero in the Fourier
transform of line profiles. Here, the high-quality Fe~{\sc i} line 
at 6252.554~\AA\ ($W_{\odot} \approx 124$~m\AA; occasionally used for 
this purpose; {\it e.g.}, Gray, 1977) was adopted as the representative test line. 
After the line depth profiles, $D({\lambda}) \equiv 1 - F_{\lambda}/F_{\rm cont}$,  
were simulated for 6 cases of different conditions ({\it cf.} Table~4),
their Fourier transforms, 
$d(\sigma) \equiv \int D(\lambda) \exp(2\pi i \lambda) {\rm d}\lambda$,
were computed as illustrated in Figure~7.\footnote{Meaningful information can also 
be read from Figure~7c, First, $q_{1}$ is only marginally influenced 
({\it i.e.}, slightly shifting to higher frequencies) by inclusion of 
macroturbulence and rotation as seen from the results of Cases 1--3. 
Second, the first zero of the rotational broadening function (so-called 
Carroll's zero) is not observed at all in the transform for Case~3, 
which means that the rotational effect on the profile is no more 
expressed by a convolution of the broadening function for the case 
of slow rotators ({\it cf.} Takeda, 2019).} 
Since the positions of the first zero ($q_{1}$)
are 0.143--0.148~km$^{-1}$s when the realistic 
$\theta$-dependent $\xi$ is used (Cases 1, 2, and 3), this corresponds to 
$\xi \approx$1.1--1.2~km~s$^{-1}$ in the conventional assumption of constant 
$\xi$, as expected from the $q_{1}$ values for Cases~0a, 0b, and 0c 
(0.170, 0.150, and 0.133~km$^{-1}$s; {\it cf.} Table~4).\footnote{It should not
be misunderstood that $\xi \approx$1.1--1.2~km~s$^{-1}$ is actually derived 
for the Sun from the line flux profile analysis. As briefly remarked 
in Section~1, $\xi$(profile) tends to be lower than $\xi$(EW); for example,
Gray (1977) derived $\xi = 0.5$~km~s$^{-1}$ from the profile analysis of
Fe~{\sc i} 6252.554 line. See, {\it e.g.}, Section~4.3 in 
Takeda {\it et al.} (1996) for a consideration on this discrepancy.}

Both of the two tests described above have suggested that $\xi$ values 
resulting from the solar disk-integrated spectrum tend to be slightly 
larger by $\sim$~20\% than that at the disk center.
This is a natural consequence in a sense, because the flux spectrum
may be regarded as roughly equivalent to the intensity spectrum at 
$\mu = \cos \theta = 2/3$  (so-called Eddington--Barbier relation),
where Equation~(1) yields $\xi = 1.18$~km~s$^{-1}$.   
Alternatively, the mean $\langle \xi \rangle$ averaged over the solar disk
may be expressed as 
\begin{equation}
\langle \xi \rangle = 
\left. \int_{0}^{\pi/2}I(\theta){\xi(\theta)}\sin\theta \cos\theta {\rm d}\theta \right/
\int_{0}^{\pi/2}I(\theta) \sin\theta \cos\theta {\rm d}\theta.
\end{equation}
Let us further write the $\theta$-dependence of the continuum specific intensity as
\begin{equation}
I(\theta) = I_{0} (1- u + u \cos\theta),
\end{equation}
where $I_{0}$ is the disk center intensity and $u$ is the limb-darkening 
coefficient (around $\approx$~0.7 for the case of the Sun at the visual 
region of $\lambda \approx$~5000--6000~\AA; {\it cf.} Fig.~17.6 in Gray, 2005).
Inserting Equations~(1) and (3) to Equation~(2), we obtain
$\langle\xi\rangle = 1.20$~km~s$^{-1}$.
Accordingly, $\xi$(flux) is expected to be slightly larger than $\xi$(center)
by $\approx$~20\% from Equation~(1). Although this inequality is not
clearly confirmed in our results in Table~3, such a tendency is observed 
in the literature data in Table~1. 

\subsection{Physical cause of center--limb variation}

Finally, it is worth briefly mentioning the physical background
regarding the systematic increase of $\xi$ towards the limb.
Historically, this trend was occasionally discussed (especially until 
1970s) in terms of its height-dependence or intrinsic anisotropy.
However, such a classical interpretation appears to be of limited 
significance, because microturbulence itself is not so much a 
physically well-defined entity as a practically useful fudge parameter. 

In any case, the former explanation is unlikely, because $\xi$ is
reported to rather decrease with height in the solar photosphere  
({\it e.g.}, Gray, 1978) from 1~km~s$^{-1}$ ($\tau_{5000} \approx 1$) to 
0.5~km~s$^{-1}$ ($\tau_{5000} \approx 10^{-3}$); thus, $\xi$ would 
rather decrease toward the limb (contrary to the observational fact) 
as the line-forming layer moves progressively higher. 
Likewise, the latter anisotropic microturbulence model does not 
seem to be relevant. If a simple anisotropic Gaussian microturbulence 
model is assumed, the $\theta$-dependence of $\xi$ would be
described as 
$\xi(\theta)^{2} = (\xi_{\rm R}\cos\theta)^2+(\xi_{\rm T}\sin\theta)^2$,
where $\xi_{\rm R}$ and $\xi_{\rm T}$ are the radial and tangential components.
which may be set as 1~km~s$^{-1}$ and 2~km~s$^{-1}$ in the present case, respectively.
The $\xi$ {\it vs.} $\theta$ relation expected in this case is shown in Figures~4a and 4b
(dashed line) for comparison, where we can see that this does not fit the observed 
trend at all.
 
The reason why more or less saturated lines show strength excess 
towards the limb (eventually causing an apparent increase in $\xi$) should 
be attributed to a more intricate process in which inhomogeneous 
velocity fields are involved in a complex manner. For example, Nordlund (1980) 
showed based on his 3D hydrodynamical calculations that, while the line 
intensification near the disk center is caused by the velocity gradients 
associated with time-dependent granular motions, the strengthening near to 
the limb is attributed rather to a geometrical effect due to the existence 
of velocity fluctuations along the line of sight penetrating more granular cells.
Accordingly, it is requisite to invoke 3D hydrodynamical model atmospheres
for understanding the nature of center--limb variations in $\xi$.  
In this context, Steffen, Caffau, and Ludwig's (2013) theoretical 
investigation on the apparent behaviors of classical micro/macro-turbulence 
based on CO5BOLD 3D model atmospheres is worthwhile to mention.
Although the detailed center--limb trends of these parameters are not 
presented, their simulations indicate $\xi$(flux)/$\xi$(center) ratios 
of $\approx$~1.2--1.3, when the results at the same attributes 
(species, equivalent widths, excitation potential, {\it etc.})
are compared with each other ({\it cf.} their Figures 1 and 3).
This extent of theoretical ratio predicted from 3D calculations 
is almost consistent with the empirical results 
(around $\approx$~1.2) discussed in Section~4.4, which is satisfying. 

\section{Summary and Conclusion}

Microturbulence ($\xi$) is an important parameter in stellar spectroscopy,
which was originally introduced by the necessity of reproducing 
the enhanced strength of saturated lines. Formally, this parameter 
is included (as if an additional thermal broadening) in the Doppler 
broadening of line opacity profile, while implicitly assuming 
that the characteristic scale-length of the relevant turbulent 
velocity field is considerably smaller than the photon mean-free-path.

Although it has been suggested that this $\xi$ at the limb tends
to be larger than that at the disk center, the detailed behavior 
regarding how $\xi$ changes on the solar disk has not been 
investigated so far. The purpose of this study was to clarify this point.

As such, the $\xi$ values were determined from the equivalent widths 
of 46 Fe~{\sc i} lines measured on the 32 points from the center to 
the limb, by following the conventional requirement that the abundances 
derived from different lines do not show any systematic dependence upon 
the line strengths, where Blackwell {\it et al.}'s (1976) procedure 
(searching for minimum abundance dispersion) was adopted in practice. 
At each point, the solution of $\xi$ could be successfully established, 
and the mean Fe abundance $\langle\log\epsilon\rangle$ was also derived 
as a by-product. 
  
The resulting $\langle\log\epsilon\rangle$ values turned out remarkably
constant irrespective of the position on the solar disk (the mean
averaged over 32 points is 7.411 with a standard deviation of 0.005),
which may suggest that the classical line formation treatment is
practically valid at any point regardless of $\theta$ (as long as 
abundance determination is concerned).

The systematic trend of increasing $\xi$ towards the limb could 
be confirmed. More precisely, the run of $\xi$ (with a probable 
error of $\approx \pm 0.2$~km~s$^{-1}$) with $\theta$ is 
only gradual from $\approx$~1.0~km~s$^{-1}$ (at $\sin\theta = 0$: disk 
center) to $\approx$~1.3~km~s$^{-1}$ (at $\sin\theta \approx  0.7$: 
two-thirds of radial distance), while increasing thereafter more steeply 
up to $\approx$~2~km~s$^{-1}$ (at $\sin\theta = 0.97$: limb).
  
Based on the derived $\xi$ {\it vs.} $\theta$ relation, test calculations 
were performed to see how much microturbulence (if assumed constant) 
would be obtained from the flux spectrum of the disk-integrated Sun.
It revealed that $\xi$(flux) is about $\approx 20$\% larger than
than $\xi$(center). This result is consistent with the prediction 
from 3D hydrodynamical calculations carried out by Steffen, 
Caffau, and Ludwig (2013).   
\newline
\newline
{\bf Disclosure of Potential Conflicts of Interest}\\
The author declares that he has no conflicts of interest.

\setcounter{table}{0}
\begin{table}[h]
\tiny
\caption{Literature data of solar microturbulence (after 1976).}
\begin{center}
\begin{tabular}
{cccccl}\hline 
Reference & Obs.part & $\mu$  & $\xi$ & Method & Remark\\
(1) & (2) & (3) & (4) & (5) & (6) \\
\hline
Blackwell {\it et al.} (1976) & center & 1.0 & 0.80 & EW &  \\
 & outer disk & 0.5 & 1.25 & EW &  \\
 & limb & 0.3 & 1.40 & EW &  \\
 & limb & 0.2 & 1.70 & EW &  \\
 & flux & --- & 1.18 & EW &  \\
Gray (1977) & flux & --- & 0.5 & profile &  \\
Holweger, Gehlsen, and Ruland (1978) & center & 1.0 & 1.0 & EW &  \\
 & limb & 0.3 & 1.6 & EW &  \\
Blackwell, Booth, and Petford (1984) & center & 1.0 & 0.85 & EW &  \\
Gadun and Kostyk (1990) & center & 1.0 & 0.89 & EW & mean of 6 values in Table~I therein \\
 & limb & 0.2 & 1.84 & EW & mean of 5 values in Table~I therein \\
 & flux & --- & 1.07 & EW & mean of 4 values in Table~I therein \\
Gadun (1994) & center & 1.0 & 0.81 & EW & result from Fe~{\sc i} lines \\
 & flux & --- & 1.11 & EW & result from Fe~{\sc i} lines \\
 & center & 1.0 & 0.86 & EW & result from Fe~{\sc ii} lines \\
 & flux & --- & 1.20 & EW & result from Fe~{\sc ii} lines \\
Takeda (1995) & flux & --- & 0.51 & profile & mean of 15 lines with Case-C solutions \\
Sheminova and Gadun (1998) & flux & --- & 0.8 & profile &  \\
Grevesse and Sauval (1999) & center & 1.0 & 0.8 & EW &  \\
Takeda, Ohkubo, and Sadakane (2002) & center & 1.0 & 0.76 & EW & use of Grevesse and Sauval's (1999) EW \\
 & flux & --- & 0.79 & EW & use of Meylan {\it et al.}'s (1993) EW \\
 & flux & --- & 0.86 & EW & use of EW measured by themselves \\
Takeda {\it et al.} (2005) & flux & --- & 0.96 & EW & Moon spectrum \\
Pavlenko {\it et al.} (2012) & flux & --- & 0.75 & EW & result from Fe~{\sc i} lines \\
 & flux & --- & 1.5 & EW & result from Fe~{\sc ii} lines \\
Sheminova (2017) & flux & --- & 0.85 & profile &  \\
Sheminova (2019) & flux & --- & 0.78 & profile &  \\
\hline
\end{tabular}
\end{center}
\tiny
(1) References. (2) Observed part of the Sun (where ``flux'' means the 
disk-integrated spectrum of Sun-as-a-star). (3) Value of direction cosine 
$\mu$ ($\equiv \cos\theta$) at the point of observation. 
(4) Method of determination, where ``EW'' is the standard method using 
the equivalent widths of spectral lines, while ``profile'' is based on
the analysis of spectral line profiles (mostly by using the Fourier technique, 
except for the usual profile fitting method adopted by Takeda, 1995). 
(5) Resulting value of microturbulence (in km~s$^{-1}$). (6) Specific remarks.
\end{table}

\setcounter{table}{1}
\begin{table}[h]
\scriptsize
\caption{Atomic and observational data of adopted 46 Fe~{\sc i} lines.}
\begin{center}
\begin{tabular}
{cccrrrrr}\hline 
$\lambda$ & $\chi_{\rm low}$ & $\log gf$  & $W_{\rm d.c.}^{\rm GS}$ & $W_{00}$ & $W_{31}$ 
& $W_{\rm flux}$ & $K_{00}$\\
(1) & (2) & (3) & (4) & (5) & (6) & (7) & (8)\\
\hline
5225.533 & 0.110 & $-$4.789 & 70.3 & 66.9 & 89.2 & 70.1 & $-$4.97\\
5247.059 & 0.087 & $-$4.946 & 65.6 & 62.4 & 81.0 & 65.2 & $-$5.52\\
5250.217 & 0.121 & $-$4.938 & 65.5 & 61.1 & 84.0 & 64.0 & $-$5.69\\
5326.145 & 3.570 & $-$2.071 & 34.5 & 30.5 & 37.7 & 33.0 & $-$5.77\\
5412.788 & 4.440 & $-$1.716 & 19.1 & 18.6 & 24.9 & 20.7 & $-$5.59\\
5600.227 & 4.260 & $-$1.420 & 36.5 & 27.6 & 32.5 & 29.6 & $-$5.24\\
5661.348 & 4.280 & $-$1.756 & 22.0 & 19.2 & 25.1 & 21.0 & $-$5.86\\
5696.093 & 4.550 & $-$1.720 & 13.4 & 9.1 & 11.5 & 12.4 & $-$6.00\\
5701.553 & 2.560 & $-$2.216 & 86.0 & 80.9 & 92.1 & 82.6 & $-$3.11\\
5705.468 & 4.300 & $-$1.355 & 39.7 & 34.9 & 39.8 & 36.9 & $-$4.64\\
5778.458 & 2.590 & $-$3.440 & 20.2 & 15.6 & 24.8 & 19.0 & $-$9.20\\
5784.661 & 3.400 & $-$2.530 & 25.4 & 22.1 & 30.1 & 24.9 & $-$7.14\\
5855.080 & 4.610 & $-$1.478 & 21.7 & 18.6 & 25.4 & 21.2 & $-$5.28\\
5909.978 & 3.210 & $-$2.587 & 30.0 & 23.4 & 29.5 & 24.8 & $-$7.27\\
5956.700 & 0.859 & $-$4.605 & 50.3 & 47.7 & 65.1 & 51.1 & $-$7.51\\
6082.715 & 2.223 & $-$3.573 & 33.8 & 27.5 & 42.4 & 33.7 & $-$8.47\\
6120.255 & 0.910 & $-$5.970 & 4.8 & 2.9 & 6.1 & 4.5 & $-$14.69\\
6151.622 & 2.176 & $-$3.299 & 48.5 & 43.1 & 60.3 & 49.0 & $-$6.57\\
6180.208 & 2.730 & $-$2.586 & 56.0 & 47.1 & 59.8 & 50.0 & $-$5.47\\
6200.320 & 2.609 & $-$2.437 & 75.4 & 70.1 & 84.1 & 70.9 & $-$3.67\\
6219.289 & 2.198 & $-$2.433 & 91.0 & 85.0 & 103.9 & 85.8 & $-$3.40\\
6232.649 & 3.650 & $-$1.223 & 88.4 & 82.7 & 92.9 & 81.8 & $-$3.01\\
6240.652 & 2.220 & $-$3.233 & 47.6 & 44.5 & 58.6 & 47.2 & $-$6.36\\
6265.141 & 2.176 & $-$2.550 & 86.7 & 84.8 & 99.5 & 83.6 & $-$3.48\\
6271.283 & 3.330 & $-$2.703 & 20.5 & 19.8 & 27.2 & 21.5 & $-$7.55\\
6297.801 & 2.223 & $-$2.740 & 74.8 & 71.8 & 85.6 & 71.3 & $-$3.91\\
6311.505 & 2.830 & $-$3.141 & 27.0 & 20.8 & 29.7 & 25.8 & $-$8.30\\
6322.694 & 2.588 & $-$2.426 & 79.2 & 68.4 & 79.1 & 73.0 & $-$3.80\\
6353.840 & 0.910 & $-$6.477 & 1.4 & 0.9 & 1.6 & 1.3 & $-$15.21\\
6481.878 & 2.279 & $-$2.984 & 64.2 & 57.6 & 65.7 & 60.6 & $-$4.97\\
6498.946 & 0.958 & $-$4.699 & 43.7 & 40.8 & 56.3 & 44.3 & $-$8.68\\
6518.374 & 2.830 & $-$2.450 & 57.0 & 50.6 & 58.3 & 52.5 & $-$5.15\\
6574.233 & 0.990 & $-$5.004 & 25.7 & 22.8 & 37.7 & 27.5 & $-$11.51\\
6581.214 & 1.480 & $-$4.680 & 17.0 & 13.7 & 22.2 & 16.3 & $-$11.64\\
6593.880 & 2.433 & $-$2.422 & 86.5 & 81.7 & 91.0 & 81.6 & $-$3.47\\
6609.119 & 2.559 & $-$2.692 & 65.3 & 60.5 & 67.5 & 61.3 & $-$4.50\\
6625.027 & 1.010 & $-$5.336 & 13.5 & 11.6 & 20.7 & 14.6 & $-$12.74\\
6667.721 & 4.580 & $-$2.112 & 8.9 & 8.0 & 9.8 & 8.8 & $-$6.10\\
6699.142 & 4.590 & $-$2.101 & 8.0 & 6.6 & 8.6 & 7.7 & $-$6.13\\
6750.160 & 2.424 & $-$2.621 & 76.2 & 71.4 & 81.0 & 72.8 & $-$3.85\\
6752.711 & 4.640 & $-$1.204 & 37.3 & 32.4 & 35.8 & 34.2 & $-$4.46\\
6793.266 & 4.080 & $-$2.326 & 12.6 & 10.3 & 13.8 & 12.0 & $-$6.98\\
6804.003 & 4.650 & $-$1.496 & 21.4 & 16.2 & 17.5 & 18.0 & $-$5.51\\
6804.277 & 4.580 & $-$1.813 & 14.9 & 11.6 & 13.5 & 13.5 & $-$5.98\\
6837.009 & 4.590 & $-$1.687 & 17.7 & 15.1 & 17.5 & 16.3 & $-$5.74\\
6854.828 & 4.590 & $-$1.926 & 12.7 & 9.9 & 11.8 & 11.4 & $-$5.84\\
\hline
\end{tabular}
\end{center}
\scriptsize
(1) Air wavelength (in \AA). (2) Lower excitation potential (in eV).
(3) Logarithm of $g$ (lower statistical weight) times $f$ (oscillator strength)
(in dex).(4) Equivalent width (in m\AA) at the disk center measured by
Grevesse and Sauval (1999). (5) Equivalent width at the disk center 
(point 00: $\mu = 1$) taken from Table~3 in TU19.
(6) Equivalent width at the limb 
(point 31: $\mu = 0.25$) taken from Table~3 in TU19.
(7) Equivalent width for the disk-integrated Sun measured on Kurucz {\it et al.}'s
(1984) solar flux spectrum in the similar manner to that adopted by TU19.
(8) Temperature sensitivity parameter calculated at the disk center 
($K_{00} \equiv {\rm d}\log W_{00}/{\rm d}\log T$; taken from Table~3 in TU19).
The data in Columns (1)--(3) were adopted from Table~1 of Grevesse and Sauval (1999).

\end{table}

\setcounter{table}{2}
\begin{table}[h]
\scriptsize
\caption{Resulting center--limb variation of microturbulence.}
\begin{center}
\begin{tabular}
{cccccccc}\hline 
No. & $\cos\theta$ & $\sin\theta$  & $\langle\log\epsilon\rangle_{0}$ &
$\sigma_{0}$ & $\xi_{0}$ &  $\delta\xi$ & $\xi_{\rm n}$ \\ 
(1) & (2) & (3) & (4) & (5) & (6) & (7) & (8)\\
\hline
00 & 1.0000 & 0.0000 & 7.401 & 0.064 & 1.13 & 0.19 & 1.00\\
01 & 0.9995 & 0.0313 & 7.409 & 0.067 & 1.10 & 0.20 & 0.97\\
02 & 0.9980 & 0.0625 & 7.399 & 0.071 & 1.08 & 0.21 & 0.96\\
03 & 0.9956 & 0.0938 & 7.401 & 0.065 & 1.10 & 0.19 & 0.97\\
04 & 0.9922 & 0.1250 & 7.408 & 0.069 & 1.13 & 0.20 & 1.00\\
05 & 0.9877 & 0.1562 & 7.409 & 0.068 & 1.15 & 0.20 & 1.02\\
06 & 0.9823 & 0.1875 & 7.415 & 0.069 & 1.12 & 0.20 & 0.99\\
07 & 0.9758 & 0.2188 & 7.413 & 0.069 & 1.14 & 0.21 & 1.01\\
08 & 0.9682 & 0.2500 & 7.416 & 0.067 & 1.12 & 0.19 & 0.99\\
09 & 0.9596 & 0.2812 & 7.414 & 0.068 & 1.13 & 0.20 & 1.00\\
10 & 0.9499 & 0.3125 & 7.414 & 0.068 & 1.12 & 0.20 & 0.99\\
11 & 0.9391 & 0.3438 & 7.417 & 0.070 & 1.12 & 0.21 & 0.99\\
12 & 0.9270 & 0.3750 & 7.413 & 0.070 & 1.15 & 0.20 & 1.02\\
13 & 0.9138 & 0.4062 & 7.413 & 0.070 & 1.17 & 0.21 & 1.04\\
14 & 0.8992 & 0.4375 & 7.413 & 0.069 & 1.15 & 0.20 & 1.02\\
15 & 0.8833 & 0.4688 & 7.414 & 0.072 & 1.16 & 0.21 & 1.03\\
16 & 0.8660 & 0.5000 & 7.414 & 0.072 & 1.18 & 0.21 & 1.04\\
17 & 0.8472 & 0.5312 & 7.416 & 0.070 & 1.18 & 0.20 & 1.04\\
18 & 0.8268 & 0.5625 & 7.417 & 0.071 & 1.20 & 0.19 & 1.06\\
19 & 0.8046 & 0.5938 & 7.416 & 0.073 & 1.21 & 0.21 & 1.07\\
20 & 0.7806 & 0.6250 & 7.415 & 0.072 & 1.23 & 0.20 & 1.09\\
21 & 0.7545 & 0.6562 & 7.409 & 0.073 & 1.27 & 0.20 & 1.12\\
22 & 0.7262 & 0.6875 & 7.410 & 0.073 & 1.29 & 0.20 & 1.14\\
23 & 0.6953 & 0.7188 & 7.410 & 0.074 & 1.31 & 0.20 & 1.16\\
24 & 0.6614 & 0.7500 & 7.411 & 0.073 & 1.34 & 0.20 & 1.19\\
25 & 0.6242 & 0.7812 & 7.412 & 0.076 & 1.38 & 0.21 & 1.22\\
26 & 0.5830 & 0.8125 & 7.415 & 0.073 & 1.41 & 0.20 & 1.25\\
27 & 0.5367 & 0.8438 & 7.411 & 0.079 & 1.50 & 0.22 & 1.33\\
28 & 0.4841 & 0.8750 & 7.408 & 0.078 & 1.56 & 0.20 & 1.38\\
29 & 0.4227 & 0.9062 & 7.403 & 0.078 & 1.71 & 0.22 & 1.51\\
30 & 0.3480 & 0.9375 & 7.407 & 0.079 & 1.89 & 0.23 & 1.67\\
31 & 0.2480 & 0.9688 & 7.421 & 0.079 & 2.10 & 0.24 & 1.86\\
\hline
\multicolumn{3}{c}{Analysis of $W_{\rm d.c.}^{\rm GS}$} & 7.513 & 0.053 & 1.08 & 0.14 &  \\
\hline
\multicolumn{3}{c}{Analysis of $W_{\rm flux}$}& 7.434 & 0.066 & 1.15 & 0.17 &  \\
\hline
\end{tabular}
\end{center}
\scriptsize
(1) Designated number of each observed point ({\it cf.} Table~1 in TU19).
(2) Value of $\cos\theta$ (direction cosine also denoted as $\mu$), where $\theta$ 
is the angle between the line of sight and the normal to the surface.
(3) Value of $\sin\theta$, which is equivalent to the concentric radius in 
unit of the disk radius. (4) Mean of $\log\epsilon$ (logarithmic number abundance
of  Fe relative to H in the usual normalization of $\log\epsilon_{\rm H} = 12$) 
at $\sigma_{0}$. (5) Standard deviation (in dex) at the minimum abundance dispersion. 
(6) Microturbulence solution (in km~s$^{-1}$) defined by the requirement of 
minimum abundance dispersion. (7) Probable error in $\xi$ ({\it cf.} Section~3.2).
(8) Adjusted microturbulence normalized as $\xi_{\rm n} \equiv \xi_{0} /1.13$,
so that $\xi_{\rm n} =1$~km~s$^{-1}$ may hold at the disk center. 
In the last two rows are also shown the related results for comparison, 
which were obtained based on $W_{\rm d.c.}^{\rm GS}$ (disk-center equivalent 
widths by Grevesse, and Sauval, 1999) and those obtained from $W_{\rm flux}$ 
(equivalent widths for the disk-integrated Sun). 
\end{table}

\setcounter{table}{3}
\begin{table}[h]
\scriptsize
\caption{Conditions and results for Fe~{\sc i} 6252.554 flux profile simulations.}
\begin{center}
\begin{tabular}
{cccccccl}\hline 
No. & $\log\epsilon$ & $\xi$ & $\zeta$  & $v_{\rm e}$ & $W_{\rm cal,flux}$ & $q_{1}$ & Figures \\ 
(1) & (2) & (3) & (4) & (5) & (6) & (7) & (8) \\
\hline
0a  &  7.70 & 0.5  & --- & ---  & 123.3 & 0.170 & Fig.~7a,b (solid, pink) \\
0b  &  7.56 & 1.0  & --- & ---  & 123.7 & 0.150 & Fig.~7a,b (solid, grey)\\
0c  &  7.33 & 1.5  & --- & ---  & 123.9 & 0.133 & Fig.~7a,b (solid, light green) \\
\hline
1   &  7.47 & $\xi(\theta)$ & --- & --- & 123.8 & 0.143 & Fig.~7a,b,c,d (dashed, blue) \\
2   &  7.47 & $\xi(\theta)$ & $\zeta(\theta)$ & --- & 123.8 & 0.145 & Fig.~7c,d (dash-dotted, green)\\
3   &  7.47 & $\xi(\theta)$ & $\zeta(\theta)$ &  1.9 & 123.8 & 0.148 & Fig.~7c,d (solid, red)\\
\hline
\end{tabular}
\end{center}
\scriptsize
(1) Case number. (2) Assigned Fe abundance, which was adjusted in advance
so that the resulting equivalent width ($W_{\rm cal,flux}$) is almost
equal to the observed value ($\approx 124$~m\AA).
(3) Microturbulence, where constant values (in km~s$^{-1}$) 
were assigned in Cases 0a--c, while $\theta$-dependent microturbulence
as described in Equation~(1) was assumed in Cases 1--3.
(4) Macroturbulence (included only for Cases 2 and 3), for which the $\theta$-dependent 
anisotropic Gaussian form described by $\zeta(\theta) = 1.5 + 1.0\sin\theta$ 
({\it cf.} Equation~(10) in Takeda, 2019) was used.
(5) Equatorial velocity of solar rotation, which was included only in Case~3
on the assumption of rigid rotation and equator-on view (inclination angle of 
$i = 90^{\circ}$).
(6) Equivalent width of the resulting flux profile (in m\AA).
(7) First zero frequency in the Fourier transform of the profile (in km$^{-1}$s)
(8) Figure panels where the Fourier transform amplitudes
of the relevant profiles are illustrated. 
\end{table}

\begin{figure} 
\centerline{\includegraphics[width=1.0\textwidth]{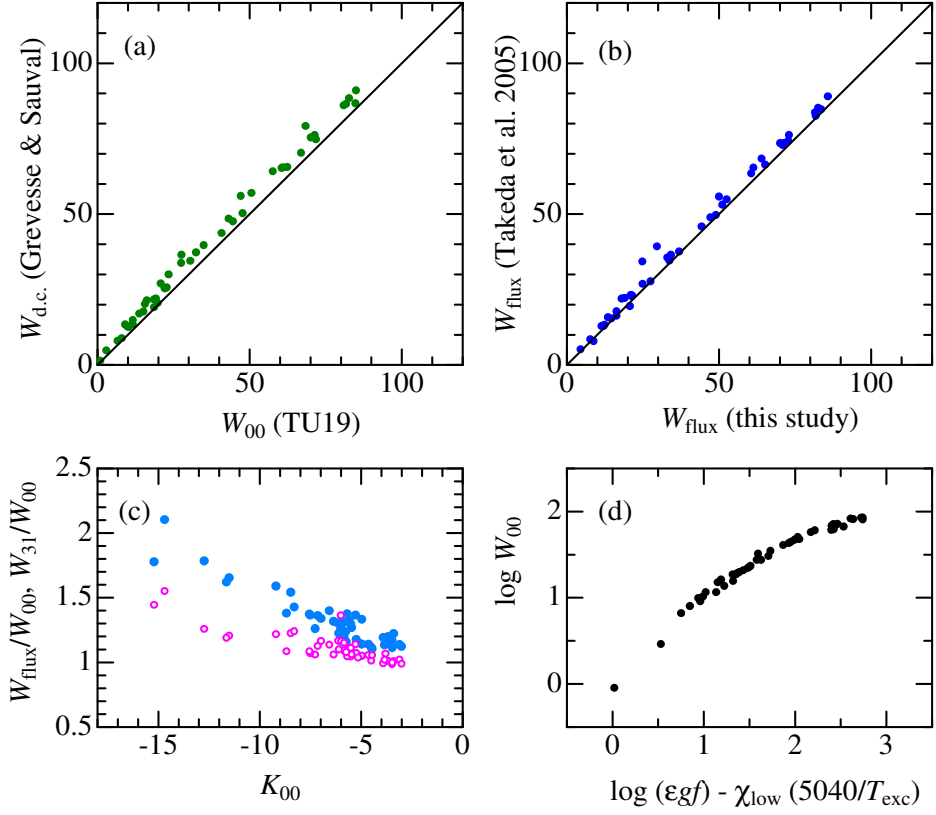}}
\caption{
(a) Comparison of Grevesse and Sauval's (1999) disk-center 
equivalent widths ($W_{\rm d.c.}^{\rm GS}$) for the Fe~{\sc i} lines 
with those of TU19 ($W_{00}$) adopted in this study.
(b) Comparison of Takeda {\it et al.}'s (2005) solar flux equivalent widths
of Fe~{\sc i} lines (measured by using the direct Gaussian fitting) 
with those newly evaluated in this study (by following the similar 
way to TU19).
(c) Limb-to-center equivalent width ratios ($W_{31}/W_{00}$; 
filled symbols) and flux-to-center ratios ($W_{\rm flux}/W_{00}$;
open symbols) plotted against $K_{00}$ (temperature sensitivity parameter). 
(d) Empirical curve of growth constructed based on the disk-center
equivalent widths ($W_{00}$) of 46 Fe~{\sc i} lines 
given in Table~2. Here, $\log W_{00}$ data are plotted against 
$\log\epsilon + \log gf - \chi_{\rm low}(5040/T_{\rm exc})$,
where $\log\epsilon$ is the abundance obtained from $W_{00}$
for $\xi = 1$~km~s$^{-1}$ and $T_{\rm exc}$ (excitation temperature) 
is assumed to be 5040~K.  
}
\end{figure}

\begin{figure} 
\centerline{\includegraphics[width=1.0\textwidth]{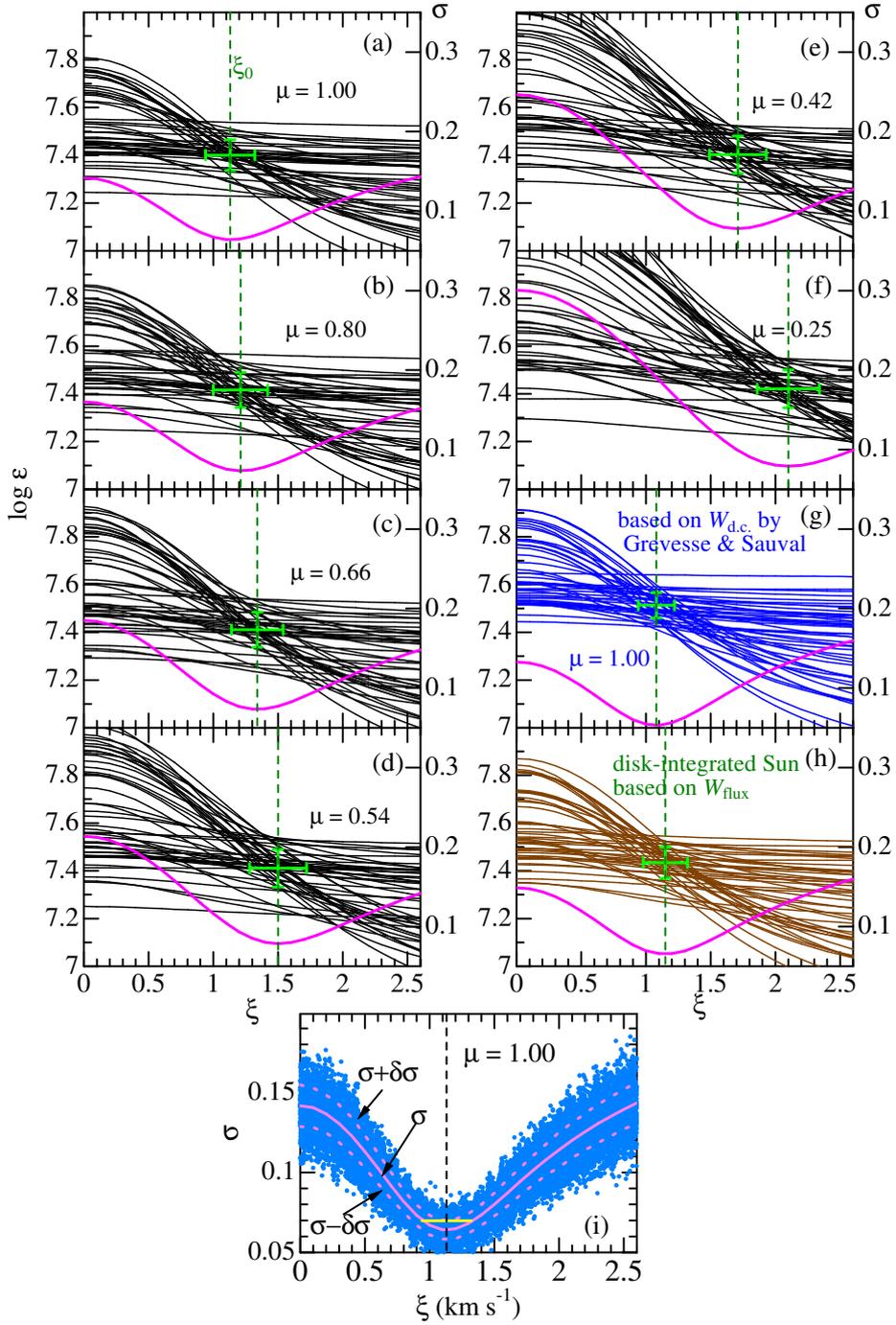}}
\caption{
Demonstrative examples of $\log\epsilon$ {\it vs.} $\xi$ diagrams for 
selected 8 cases: (a) $\mu =1.00$, (b) $\mu = 0.80$, (c) $\mu = 0.66$,
(d) $\mu = 0.54$, (e) $\mu = 0.42$, (f) $\mu = 0.25$, (g) $\mu = 1.00$ 
based on Grevesse and Sauval's (1999) $W_{\rm d.c.}^{\rm GS}$ data,
and (h) use of $W_{\rm flux}$ for the disk-integrated Sun.
In each panel, $\sigma$ (standard deviation of $\log \epsilon$) is 
plotted against $\xi$ by the thick (pink) solid line (its scale is marked 
in the right axis), and the solution 
of $\xi_{0}$ (corresponding to $\sigma_{0}$: $\sigma$ minimum) is indicated 
by the vertical dashed line. The green error bars shown at this solution point 
($\xi_{0}$, $\langle\log\epsilon\rangle_{0}$) are $\pm\delta\xi$ (probable 
error in $\xi$) and $\pm\sigma_{0}$ (dispersion of $\log\epsilon$). 
Panel (i) shows the numerical experiment for evaluation of $\delta\xi$, 
done for the case of $\mu=1.00$ corresponding to panel (a).
Here, randomly generated $(\sigma'_{k}/\sqrt{2}, k = 1, 2, \ldots, 1000)$
at each $\xi$ ({\it cf.} Section~3.2) are plotted by dots, while their mean 
(naturally equal to the original $\sigma$) and its standard 
deviation ($\pm\delta\sigma$) are depicted in solid and dashed lines,
respectively. The horizontal yellow bar shows the resulting 
extent of $\pm\delta\xi$ (see the explanation in Section~3.2 for more
details). 
}
\end{figure}

\begin{figure} 
\centerline{\includegraphics[width=1.0\textwidth]{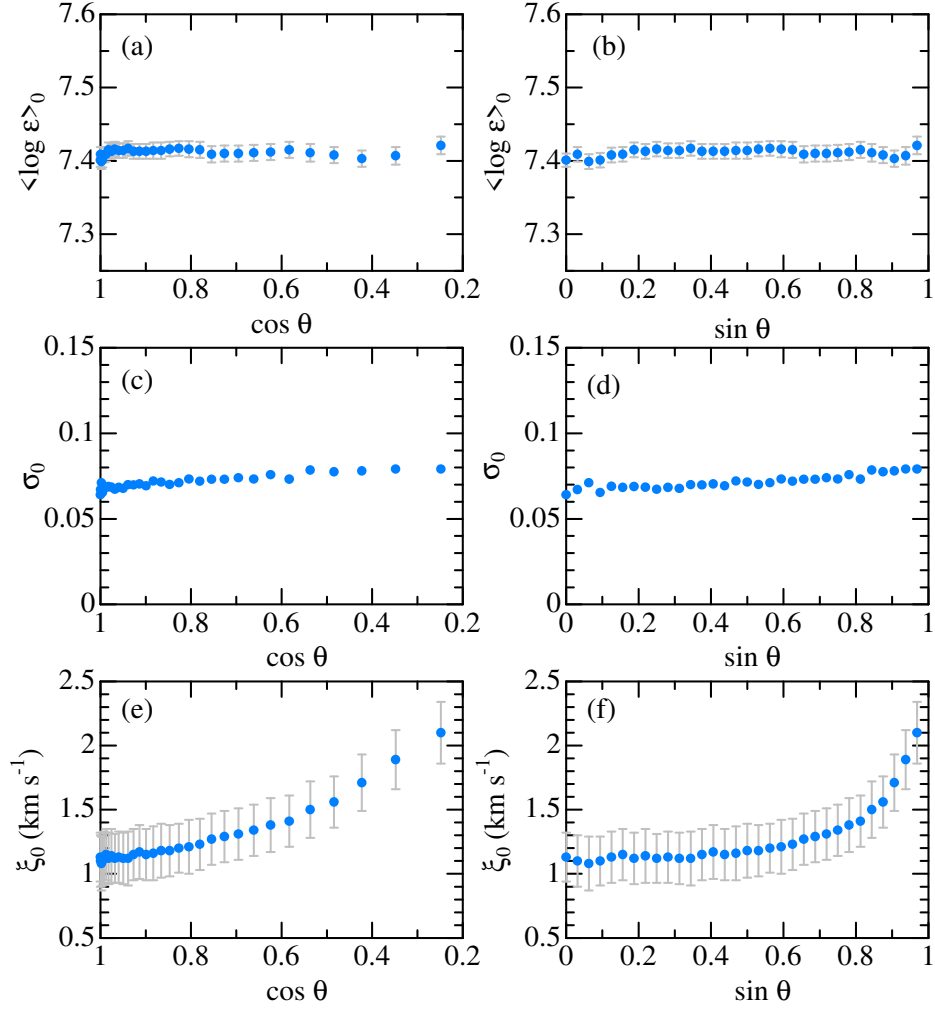}}
\caption{
Center--limb variations of $\langle \log \epsilon \rangle_{0}$ (top),
$\sigma_{0}$ (middle), and $\xi_{0}$ (bottom), which were determined 
from the $\log\epsilon$ {\it vs.} $\xi$ diagrams at each of the 33 points
on the solar disk, The results are plotted against $\cos\theta (\equiv \mu)$
and $\sin\theta$ in the left- and right-hand panels, respectively.  
The error bars attached to $\langle \log \epsilon \rangle_{0}$ (top panels) 
denote mean errors ($\pm \sigma_{0}/\sqrt{N}$, where $N=46$), while
those to $\xi_{0}$ (bottom panels) are $\pm \delta \xi$ (probable errors).
}
\end{figure}

\begin{figure} 
\centerline{\includegraphics[width=1.0\textwidth]{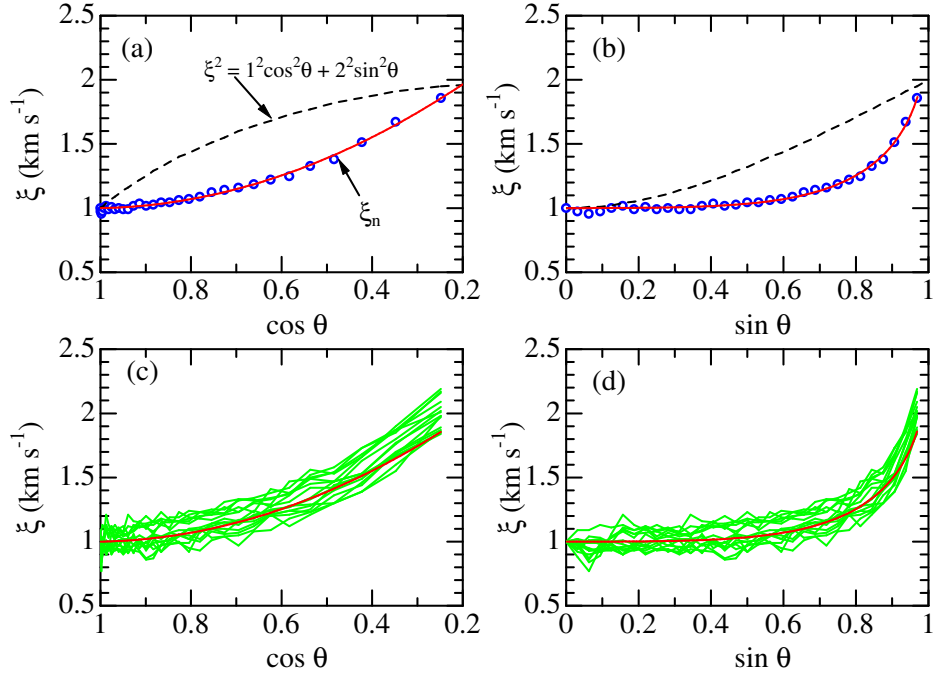}}
\caption{
Upper panels: Open symbols illustrate the adjusted angle-dependent 
solar microturbulence ($\xi_{\rm n}$: so normalized as 
to be 1~km~s$^{-1}$ at the disk center), and the best-fit curve 
(expressed by a quadratic formula: {\it cf.} Equation~(1)) is drawn by the 
solid line. The expected relation for the case of anisotropic 
Gaussian distribution ({\it cf.} Section~4.5) is also shown by 
the dashed line for comparison.
Lower panels: $\xi(\theta)$ relations for each of the 15 lines
of medium-to-large strengths ($W_{00} \ge 50$~m\AA) are overplotted 
by green lines, which were obtained by the requirement
of $\log\epsilon (\theta) = \log\epsilon_{1}$, where $\log\epsilon_{1}$
is the abundance at the disk center derived from $W_{00}$ 
by assuming $\xi = 1$~km~s$^{-1}$ and $\log\epsilon (\theta)$ is the
abundance corresponding to $W(\theta)$ (at the off-center point of 
angle $\theta$). As in the upper panels, the red line is the best-fit 
curve of $\xi_{\rm n}$. 
}
\end{figure}

\begin{figure} 
\centerline{\includegraphics[width=1.0\textwidth]{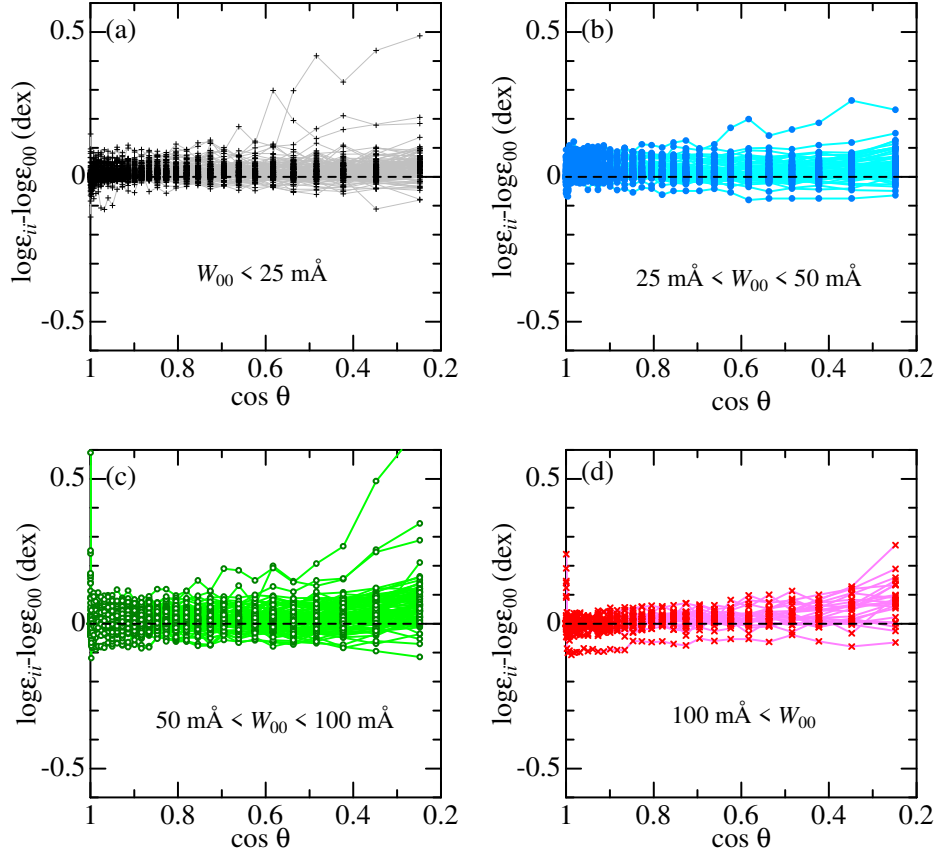}}
\caption{
Differences of abundances relative to the disk-center value
($\log\epsilon - \log\epsilon_{00}$) plotted against $\cos\theta$, 
which were derived based on the solar center--limb equivalent widths 
($W$) of 280 Fe~{\sc i} lines published in TU19 by assuming a 
$\mu$-dependent microturbulence of Equation~(1).
Each of the panels (a), (b), (c), and (d) correspond to four 
line-strength classes, which were grouped according to the 
disk-center equivalent width ($W_{00}$):
(a) $\cdots$ $W_{00} <$~25~m\AA\ (black symbols), 
(b) $\cdots$ 25~m\AA~$\le W_{00} <$~50~m\AA\ (blue symbols),
(c) $\cdots$ 50~m\AA~$\le W_{00} <$~100~m\AA\ (green symbols), and
(d) $\cdots$ 100~m\AA~$\le W_{00}$ (red symbols).
These figures should be compared to Figures~10a--d of Takeda (2019),
which show the results derived by assuming a 
constant $\xi$ of 1~km~s$^{-1}$.
}
\end{figure}

\begin{figure} 
\centerline{\includegraphics[width=1.0\textwidth]{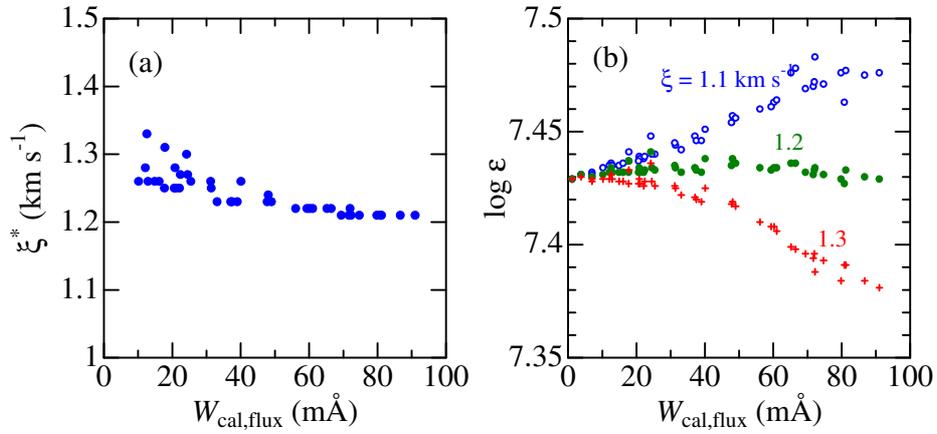}}
\caption{
(a) $\xi^{*}$ {\it vs.} $W_{\rm cal,flux}$ diagram for the 42 lines
satisfying $W_{\rm cal,flux} \ge 10$~m\AA. Here, $W_{\rm cal,flux}$
is the flux equivalent width simulated by the disk-integration 
method while assuming $\log\epsilon = 7.43$ and $\xi(\theta)$ (Equation~(1)),
and $\xi^{*}$ is the value of microturbulence necessary to reproduce 
$\log \epsilon = 7.43$ if $W_{\rm cal,flux}$ is analyzed in the 
conventional manner using a constant ({\it i.e.}, position-independent) $\xi$. 
(b) $\log\epsilon$ {\it vs.} $W_{\rm cal,flux}$ diagram for all the 46 lines,
where $\log\epsilon$ is the abundance derived by analyzing 
$W_{\rm cal,flux}$in the conventional way by using three $xi$ values of 
1.1 (open symbols), 1.2 (filled symbols), and 1.3~km~s$^{-1}$ (crosses).  
}
\end{figure}

\begin{figure} 
\centerline{\includegraphics[width=1.0\textwidth]{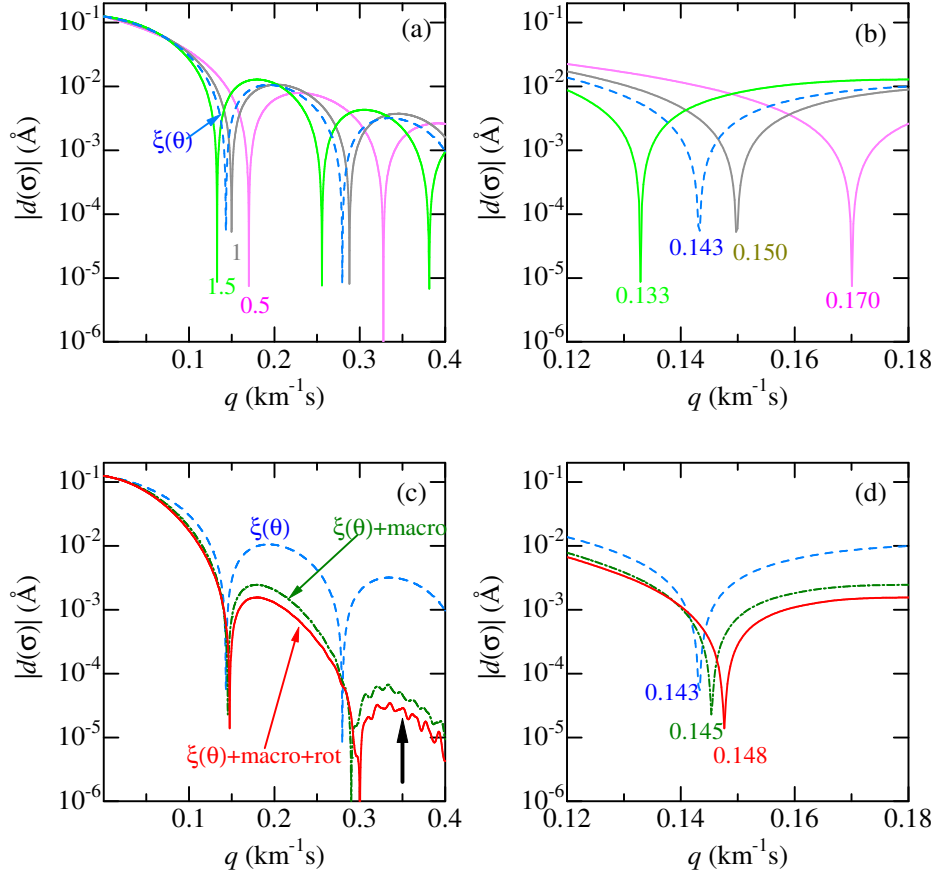}}
\caption{
Fourier transform amplitude of the flux profiles of
the Fe~{\sc i} 6252.554 line simulated for several test calculations 
(carried out according to the conditions summarized in Table~4),
plotted against wavelength-independent Fourier frequency 
($q \equiv \sigma \lambda/c$ in unit of km$^{-1}$s, where $\sigma$ 
is the usual Fourier frequency, $\lambda$ is the wavelength, and $c$ is 
the velocity of light). In the upper panels (a, b), the results for Cases~0a--c 
(cases for three constant microturbulences of 0.5, 1.0, and 1.5~km~s$^{-1}$) 
and that for Case~1 ($\theta$-dependent $\xi$)
are shown in solid and dashed lines, respectively.
The lower panels (c, d) present the results for the three Cases (1, 2, and 3)
in order to illustrate the combined effects of ($\theta$-dependent) microturbulence,
macroturbulence, and rotation. See the caption of Table~4 for more details.
The thick upward arrow in panel (c) indicates the position of
the first zero (0.35~km$^{-1}$s) corresponding to the classical 
rotational broadening function (for $v_{\rm e}\sin i = 1.9$~km~s$^{-1}$
and the limb-darkening coefficient of $\epsilon \approx$~0.6--0.7).
The right-hand panels (b, d) are essentially the same as the 
corresponding left-hand panels (a, c), except that the region of Fourier 
frequencies around the first zeros (their values are indicated) is 
expanded for clarity.  
}
\end{figure}

\end{article} 
\end{document}